\documentclass[aps,prl,twocolumn,groupedaddress,showpacs]{revtex4}

\usepackage{graphics}

\begin{document}


\title{Effective potentials for atom-atom interaction at low temperatures}


\author{Bo Gao}
\email[]{bgao@physics.utoledo.edu}
\homepage[]{http://bgaowww.physics.utoledo.edu}
\affiliation{Shanghai Institute of Optics and Fine Mechanics, 
	Chinese Academy of Sciences,
	Shanghai 201800, China}
\affiliation{Department of Physics and Astronomy,
	University of Toledo,
	Toledo, Ohio 43606}


\date{August 8, 2002}

\begin{abstract}

We discuss the concept and design of effective 
atom-atom potentials that accurately describe any physical processes 
involving only states around the threshold.
The existence of such potentials gives hope to
a quantitative, and systematic, understanding of quantum few-atom 
and quantum many-atom systems at relatively low temperatures.

\end{abstract}


\maketitle

The concept of model potential has played an important
role in many branches of physics.  
The well known examples include the Morse potential \cite{mor29}
and the Lennard-Jones potential for molecular systems, and
the hard-sphere potential and the delta-function 
pseudopotential \cite{hua57} widely used in many-body
theories, including theories for Bose-Einstein 
condensates (BEC) \cite{leg01}.

There are many reasons why one uses a model potential instead
of the ``real'' potential. Here, we only emphasize that
a problem can simply become unmanageable
if the ``real'' potential is used. For quantum systems, this
statement quickly becomes true for three or more atoms.
This explains, in part, that despite its many limitations, 
it has proven difficult to go substantially beyond the 
Gross-Pitaveskii theory for BEC that is based on the 
delta-function pseudopotential \cite{leg01}.  

Our goal here is to discuss the concept and design of model 
potentials that better reflect the reality of atom-atom interaction
than either the hard-sphere potential or the delta-function pseudopotential, 
yet simple enough to allow for applications in 
quantum few-atom and quantum many-atom systems. 
Here, being ``simple'', to a large extent, means being shallow,
as it is the depth of a potential, which can be 
measured by the number of bound states it supports, that determines
the complexity of the resulting quantum few-atom and quantum
many-atom problems.

One of our key conclusions is the following. For N-atom states around 
the N-atom threshold (such as the BEC state \cite{leg01}),
or physical processes that involve only states around the threshold
(such as the three-body recombination process \cite{nie01,sun02}),
the interaction potential between a pair of atoms, no matter how deep
it might be, can be replaced by an effective potential supporting
only one or a few bound states. Furthermore, because different partial
waves are described by the same effective potential, it can be used in
precisely the same manner as any ``real'' potential. 
And in doing so, one reduces the complexity of the resulting quantum 
few-atom and quantum many-atom problems to a level comparable to that 
for He, a level that we are quickly learning to deal 
with \cite{blu00,sun02}.

If one thinks of the delta-function pseudopotential \cite{hua57} 
as describing
the atomic interaction at the longest length scale in the 
zero-energy limit, $2\pi/k$, the natural next step is the description 
of atomic interaction at the next, shorter, length scale. 
This scale is $\beta_n=(2\mu C_n/\hbar^2)^{1/(n-2)}$, 
which characterizes the
long-range atomic interaction of the form of $-C_n/r^n$ ($n>2$).
The angular-momentum-insensitive quantum-defect 
theory (AQDT) \cite{gao01,gao00} provides a systematic
description of atom-atom interaction at this scale, and is
the basis of our discussion. 

Reference \cite{gao01} focused on two-atom systems with 
$V(r)\rightarrow -C_6/r^6$, 
but the same concepts and formulation are readily generalized to any $n>2$. 
In this general formulation, a two-atom system with an asymptotic
potential of the form of $-C_n/r^n$ ($n>2$) is described
by a dimensionless K matrix $K^c(\epsilon,l)$ and a set of
universal functions that are determined from the solutions of
\begin{equation}
\left[\frac{d^2}{dr^2} - \frac{l(l+1)}{r^2}
	+ \frac{\beta_n^{n-2}}{r^n} + \bar{\epsilon}\right]
	u_{\epsilon l}(r) = 0 ,
\label{eq:invrnsch}
\end{equation}
where $\bar{\epsilon}\equiv 2\mu \epsilon/\hbar^2$.
Specifically, $K^c(\epsilon,l)$ is defined by writing the
wave function at large distances as a linear superposition of
a pair of reference solutions of Eq.~(\ref{eq:invrnsch}):
\begin{equation}
u_{\epsilon l}(r) = A_{\epsilon l}[f^c_{\epsilon l}(r) 
	- K^c(\epsilon,l) g^c_{\epsilon l}(r)]\;,
\label{eq:wfn}
\end{equation}
where $f^c$ and $g^c$ are purposely chosen to have
the behavior
\begin{eqnarray}
f^c_{\epsilon l} &\stackrel{r\ll\beta_n}{\longrightarrow}& 
	(2/\pi)^{1/2}(r/\beta_n)^{n/4}
	\cos\left(y-\pi/4 \right) \;, 
\label{eq:fcasy0}\\
g^c_{\epsilon l} &\stackrel{r\ll\beta_n}{\longrightarrow}& 
	-(2/\pi)^{1/2}(r/\beta_n)^{n/4}
	\sin\left(y -\pi/4 \right) \;,
\label{eq:gcasy0}
\end{eqnarray}
for all energies \cite{gao01}. Here $y=[2/(n-2)](\beta_n/r)^{(n-2)/2}$. 

AQDT asserts that $K^c(\epsilon,l)$ is approximately a constant
that is independent of both $\epsilon$ and $l$,
provided that $\beta_n$ is greater than other, energy-independent,
length scales present in the system \cite{gao01}.
For our purposes here, the most important conclusion of AQDT is
the following.
\textit{To the extent of $K^c(\epsilon,l)$ being energy and 
angular-momentum independent, potentials with the same 
type of long-range behavior (namely the same $n$) and the same 
$K^c$ have, on a scaled energy basis, the same bound spectra 
and scattering properties around the threshold} \cite{gao01}. 
Here the bound spectra and scattering properties include
all angular momentum states for which $K^c$ remains 
approximately $l$-independent.

Ignoring scaling relations implied by this statement,
which will discussed elsewhere \cite{gaoupb},
AQDT gives the following simple prescription for designing an effective
potential that has the same physical properties around the threshold as the
system of interest. First, choose a model potential. The
only restriction is that it should have the right asymptotic behavior.
Second, adjust the short range parameters of the model potential
so that
\begin{equation}
K^c_{\rm mod}(0,l) = K^c(0,l) \;,
\label{eq:c1}
\end{equation}
for one particular $l$. These two conditions do not uniquely determine
an effective potential. Another auxiliary condition, which gives a
convenient characterization of the depth of a potential, is the number
of bound levels supported by a model potential for a particular $l$, $N_{l}$.
For classes of model potentials discussed below, these conditions
uniquely determine a model potential.

We stress here two classes of model potentials for which $K^c(0,l)$, and
the number of bound levels for each $l$, $N_l$, can be found
analytically. 
One class is of the type of a hard-sphere with
an attractive tail:
\begin{equation}
V_{HST}(r) = \left\{ \begin{array}{lcl}
	\infty & , & r \le r_0 \\
	-C_n/r^n & , & r>r_0 \end{array} \right. \;,
\label{eq:HST}
\end{equation}
which will be denoted by HST.
The other class is of the type of Lennard-Jones LJ$(n,2n-2)$: 
\begin{equation}
V_{LJn}(r) = -C_n/r^n+C_{2n-2}/r^{2n-2} \;,
\label{eq:LJn}
\end{equation}
which will be denoted by LJn. In particular, this potential corresponds to
a LJ(6,10) potential for $n=6$.

For HST potentials, it is not difficult to show that the $K^c$ parameter at
zero energy is given by \cite{gaoupb}
\begin{equation}
K_{\text{HST}}^c(0,l)
	= -\frac{J_{\nu_0}(y_0)\cos(\pi\nu_0/2)-Y_{\nu_0}(y_0)\sin(\pi\nu_0/2)}
		{J_{\nu_0}(y_0)\sin(\pi\nu_0/2)+Y_{\nu_0}(y_0)\cos(\pi\nu_0/2)} ,
\end{equation}
where $\nu_0=(2l+1)/(n-2)$, $J$ and
$Y$ are the Bessel functions \cite{abr64}, and $y_0=[2/(n-2)](\beta_n/r_0)^{(n-2)/2}$.
The number of bound levels for angular momentum $l$
is given by  
\begin{equation}
N_{\text{HST}}(l) = \left\{ \begin{array}{lcl}
	m & , & j_{\nu_0,m} \le y_0 < j_{\nu_0,m+1} \\
	0 & , & y_0 < j_{\nu_0,1}
	\end{array} \right.\;,
\end{equation}
where $j_{\nu_0,m}$ ($m\ge 1$), is the m-th zero of the 
Bessel function $J_{\nu_0}(x)$ \cite{abr64}.

For LJn potentials, the following results can be derived with the help of 
a local scaling transformation \cite{gaoupb}.
\begin{equation}
K_{\text{LJn}}^c(0,l) = \tan(\pi\nu_0/2)
	[1+h_l(z_0)][1-h_l(z_0)]^{-1} ,
\end{equation}
where $z_0=(\beta_n/\beta_{2n-2})^{n-2}/[2(n-2)]$,
\begin{equation}
h_l(z_0) = z_0^{\nu_0}\frac{\sin\pi(z_0+1/2-\nu_0/2)\Gamma(z_0+1/2-\nu_0/2)}
	{\sin\pi(z_0+1/2+\nu_0/2)\Gamma(z_0+1/2+\nu_0/2)} ,
\end{equation}
and $\nu_0=(2l+1)/(n-2)$.
The number of bound levels for any $l$ is given by
\begin{equation}
N_{\text{LJn}}(l) = \left\{ \begin{array}{lcl}
	\left[z_0+\frac{1}{2}-\frac{\nu_0}{2}\right] & , & z_0\ge (\nu_0+1)/2 \\
	0 & , &  z_0<(\nu_0+1)/2 
	\end{array} \right. ,
\end{equation}
where $[x]$ means the greatest integer less than or equal to $x$.
We note that scattering lengths, which are much more restrictive 
concepts \cite{lev63,gao00}, can be derived from $K^c(0,l)$. 
For example, the $s$ wave scattering length,
which is defined only for $n>3$, can be obtained from $K^c(0,l=0)$:
\begin{equation}
a_{l=0}/\beta_n = \left[b^{2b}\frac{\Gamma(1-b)}{\Gamma(1+b)}\right]
	\frac{K^c(0,0) + \tan(\pi b/2)}{K^c(0,0) - \tan(\pi b/2)} \;,
\label{eq:a0}
\end{equation}
where $b=1/(n-2)$.

From these results, the HST and LJn types of potentials can be readily 
designed according to Eq.~(\ref{eq:c1}) to have the desired 
$K^c(0,l)$, and the desired $N_l$.
Table~\ref{tb:NaEffPot} gives a selected set of designs for 
the triplet state of a $^{23}$Na dimer.
Here the potentials are designed to have $K^c(0,l=0) = 13.57$,
which is found numerically using the latest potential
for a sodium dimer \cite{lau02}. From Eq.~(\ref{eq:a0}), this corresponds 
to an $s$ wave scattering length of 64.57 a.u. The number of bound $s$
wave levels supported by this ``real'' potential is found numerically 
to be 16.

\begin{table}
\caption{Selected data, all in atomic units, for effective potentials designed for the
triplet state of a $^{23}$Na dimer. All potentials have $C_6 = 1556$ a.u. \cite{der99}
and $K^c(0,l=0)=13.57$. $N_{l=0}$ is the number of $s$ wave bound levels supported by 
the potential. $D_e$ is a derived parameter presented for discussion.
\label{tb:NaEffPot}}
\begin{ruledtabular}
\begin{tabular}{c|cc|cc}
 & \multicolumn{2}{c|}{LJ(6,10)} & \multicolumn{2}{c}{HST} \\
\hline
$N_{l=0}$ &  $C_{10}$   & $D_e$ & $r_0$  & $D_e$ \\
\hline
1  & 1.65080e+9 & 2.64709e-7 & 3.22715e+1 & 1.37752e-6 \\
2  & 4.05415e+8 & 2.17501e-6 & 2.40025e+1 & 8.13720e-6 \\
4  & 1.00245e+8 & 1.76896e-5 & 1.74332e+1 & 5.54309e-5 \\
16 & 6.21105e+6 & 1.14700e-3 & 8.89948e+0 & 3.13200e-3 \\
\end{tabular}
\end{ruledtabular}
\end{table}

\begin{figure}
\scalebox{0.4}{\includegraphics{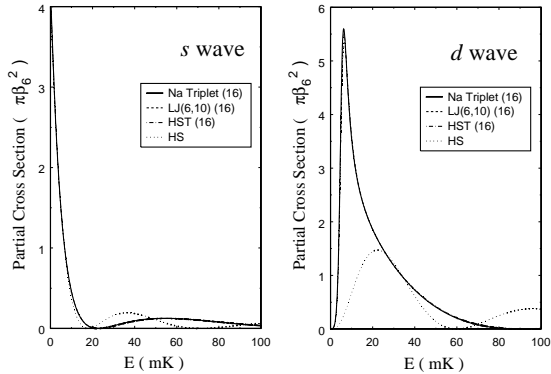}}
\caption{Comparison of $s$ and $d$ wave partial cross sections of effective LJ(6,10)
and HST potentials, both designed to support sixteen $s$ wave bound levels and have
$K^c(0,l=0)=13.57$, with the ``real'' Na-Na partial cross
sections \cite{Na}. The number in the parenthesis represents the number of $s$ 
wave bound levels supported by a potential.
Results for a hard-sphere (HS) potential are also shown for comparison.
\label{Figure1}}
\end{figure}

Figure~\ref{Figure1} shows the comparison of $s$ and $d$ wave partial cross sections 
of effective LJ(6,10) and effective HST potentials, both designed to support 
16 $s$ wave bound levels and have a $K^c(0,l=0) = 13.57$, with the Na-Na
partial cross sections computed from the ``real'' potential \cite{lau02}.
The results are hardly distinguishable over a wide range of energies. 
In comparison, the hard-sphere potential (HS) fails 
quickly away from the threshold for the $s$ wave, and gives completely wrong results 
for the $d$ wave.

\begin{figure}
\scalebox{0.4}{\includegraphics{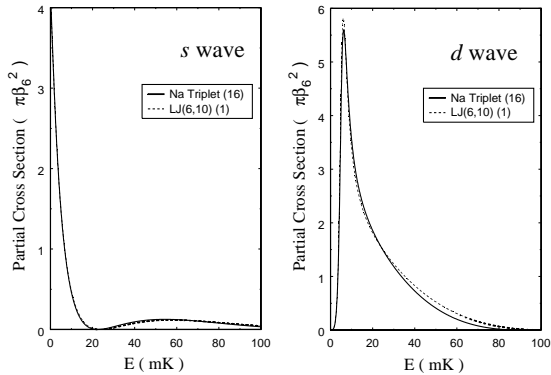}}
\caption{Comparison of $s$ and $d$ wave partial cross sections for an 
effective LJ(6,10) potential, designed to support a single $s$ wave 
bound level, with the ``real'' Na-Na cross sections \cite{Na}. \label{Figure2}}
\end{figure}

This result confirms the concept of effective potential based on AQDT.
For our purposes here, however, what is more important is how shallow the effective 
potentials can be while still maintain a good description of low-energy 
characteristics of a real system.
Figure~\ref{Figure2} shows the comparison of $s$ and $d$ wave partial cross sections 
of an effective LJ(6,10) potential, in this case designed to support only a single $s$ 
wave bound state, with the ``real'' Na-Na results \cite{Na}. The agreements 
remain excellent.

\begin{figure}
\scalebox{0.4}{\includegraphics{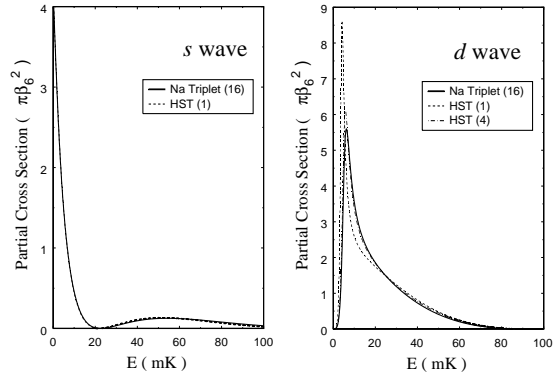}}
\caption{Comparison of $s$ and $d$ wave partial cross sections for effective HST
potentials, designed to support one and four $s$ wave bound levels, 
with the ``real'' Na-Na cross sections \cite{Na}. \label{Figure3}}
\end{figure}

Figure~\ref{Figure3} shows similar results for a HST potential. In this case, 
the HST potential that supports only a single $s$ wave bound state does not do as
well near the $d$ wave shape resonance. But it quickly improves, monotonically, as the
number of bound levels it is designed to support increases. By $N_{l=0}=4$, a good 
agreement is achieved.

\begin{figure}
\scalebox{0.4}{\includegraphics{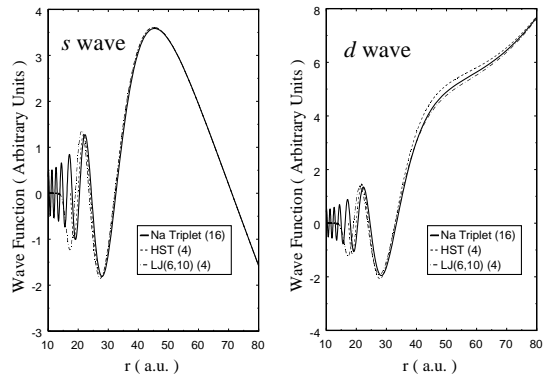}}
\caption{Zero-energy wave functions of effective potentials 
with four $s$ wave bound levels compared with the ``real'' Na-Na
wave functions \cite{Na}. \label{Figure4}}
\end{figure}

The robustness of these designs is not limited to the description of scattering
properties, it also applies to the energies of bound states that are close
to the dissociation threshold, and to the wave functions. 
For example, for effective potentials supporting
four $s$ wave bound levels, the HST gives a binding energy of 0.2027 GHz for the
least-bound $s$ state, the LJ(6,10) gives 0.2003 GHz. Both in good agreement with
the result for the ``real'' potential, which gives 0.2044 GHz \cite{Na}.
Figure~\ref{Figure4} shows that for effective potentials supporting four $s$ wave
bound levels, the wave functions are well represented down to $r=20$ a.u., 
covering basically all regions of space in which there is an appreciable amplitude.

We stress that while only the results for sodium are presented here.
They are used to illustrate much more general concepts. 
As the number of bound levels supported by an effective potential increases,
all physical properties of states around the threshold
converge to the same results (see Figs.~\ref{Figure1}-\ref{Figure4}). 
This is the \textit{shape-independence at 
length scale $\beta_n$}. The converged results, properly
scaled, represent a set of universal properties shared by all quantum 
systems with the same type of long-range potential, and characterized 
by the same, $l$-independent constant 
$K^c = \lim_{\beta_{x}/\beta_n\rightarrow 0} K^c(0,l)$ \cite{gao01},
where $\beta_{x}$ represent the next shorter length scale present in the
system. The examples presented here, Figs.~\ref{Figure2}-\ref{Figure4}, 
show how quickly this set of universal properties
are approached as one increases the number of bound levels
supported by an effective potential. 
This quick convergence is due to the fact that deviations
from the universal behavior depend on a high power of 
$\beta_{x}/\beta_n$ \cite{gaoupb}. 
Other quantum systems differ from Na primarily in $K^c$ \cite{gao01}, 
which does not effect this rapid convergence. 

Note that we did not make any distinction between two-atom and N-atom 
quantum systems in the statements above, because the same apply to a 
N-atom system. A short argument
is simply that diffuse states, in which atoms are mostly at large distances
relative to each other, only couple coherently to other diffuse states.
A longer argument can proceed as follows. The correct $K^c(0,0)$, 
and therefore $a_{l=0}$, ensures the correct
results at the mean-field level \cite{leg01}. The correct two-atom
wave function ensures the correct two-atom correlation. It also ensures 
the correct three-atom correlation, as follows. Think of a three-atom
as a two-atom perturbed by another. Frank-Condon considerations tell one that
only two-atom states around the threshold are significantly coupled.
This means by having the correct two-atom wave functions around the
threshold, one has also the correct three-atom correlation around the
threshold. \dots

To illustrate the savings of computer resources as a result
of using an effective potential, consider the problem of N interacting
atoms in a symmetric trap of frequency $\omega$.
If a real potential is used, the fact we need
to represent the length scale of $(2\mu D_e/\hbar^2)^{-1/2}$ 
means we need roughly
$D_e/\hbar\omega$ number of harmonic oscillator states for each atom, for a
total of $(D_e/\hbar\omega)^N$ number of states (ignoring statistics). 
If an effective potential
is used, the corresponding number is $(D_{e,\text{eff}}/\hbar\omega)^N$.
Thus the saving in the size of basis set is characterized by the factor
$(D_{e,\text{eff}}/D_e)^{N}$. For the triplet state of Na, 
$D_{e,\text{eff}}/D_e$ is of the order of $10^{-3}$
if an effective potential with two $s$ wave bound levels is used 
(see Table~\ref{tb:NaEffPot}). 
This corresponds to a saving in the size of basis set of $10^{9}$ fold 
just for a three-atom problem. Even greater savings are achieved for deeper
potentials or for more atoms.
From another angle, for effective potentials with $N_{l=0}\sim 1$, 
all length scales shorter than $\beta_n$ have effectively been eliminated.
It is this elimination of short length scales that makes a complex problem
more manageable. 

On the other hand, if a good description over an even wider range of energies 
around the threshold is desired, the same methodology can be carried 
to scales shorter than $\beta_n$ (e.g., $\beta_8$ for atoms with a 
$-C_6/r^6$ long-range interaction and a $-C_8/r^8$ correction) \cite{gaoupb}. 
However, because the ratio $\beta_8/\beta_6$ is different for different 
systems, the results become dependent upon one more system-specific parameter. 
At this stage, going to shorter length scale seems useful only in 
specialized two-atom applications \cite{gaoupb}. We also point out that
it is around the threshold that the quantum effects are most important
\cite{gao99b}.

In conclusion, we have established the concept and design of effective 
potentials describing atomic interaction at the length scale of $\beta_n$. 
It is the scale that one has to deal with in studying quantum few-atom 
\cite{nie01} and quantum many-atom systems \cite{leg01} at finite 
temperature, of high density, or under strong confinement. We expect the 
method presented to play a role in our understanding of some of the more
complex systems and processes at low temperatures, such as the
three-body recombination process \cite{nie01,sun02}, 
excited clusters states \cite{blu00}, and quantum liquids \cite{leg01}. 
In all cases, one can look for, and verify universal properties at the 
scale of $\beta_n$, by comparing results from different designs, 
such as HST and LJn, and by checking convergences as one relaxes an 
effective potential towards more bound levels.

\begin{acknowledgments}
I thank Eite Tiesinga, Mike Cavagnero, and Brett Esry for helpful discussions.
Special thanks goes to Eite for providing the potential for sodium dimer.
This work was supported in part by the National Natural Science Foundation of 
China (No. 19834060) and the Key Project of Knowledge Innovation Program of 
Chinese Academy of Science (No. KJCX2-W7),
and in part by the US National Science Foundation.

\end{acknowledgments}

\bibliography{effpotl}

\begin{thebibliography}{15}
\bibitem[{Na()}]{Na}
\bibinfo{note}{All results for Na dimer are computed from the potential of
  \cite{lau02}.}
\end{thebibliography}

\end{document}